\documentclass[prb,twocolumn,showpacs,floatfix]{revtex4}
\usepackage{epsfig,color}
\usepackage{amsmath}

\begin{document}
\title{Carrier-carrier inelastic scattering events for spatially separated electrons: magnetic asymmetry and turnstile electron transfer}
\author{M.R. Poniedzia{\l}ek and B. Szafran}
 \address{AGH University of Science and Technology, \\ Faculty of Physics and Applied Computer Science,
al. Mickiewicza 30, 30-059 Krak\'ow, Poland}
\date{\today}

\begin{abstract}
We consider a single electron traveling along  a strictly one-dimensional quantum wire
interacting with another electron in a quantum ring capacitively coupled to the wire.
We develop an exact numerical method for treating the scattering problem within the stationary two-electron wave function
picture. The considered process conserves the total energy but the electron within the wire
passes a part of its energy to the ring. We demonstrate that the inelastic scattering
results in both magnetic asymmetry of the transfer probability and a turnstile action of
the ring on the electrons traveling separately along the ring.
We demonstrate that the inelastic backscattering and / or inelastic electron transfer can
be selectively eliminated from the process by inclusion of an energy filter into the wire
in form of a double barrier system with the resonant energy level tuned to the energy
of the incident electron. We demonstrate that the magnetic symmetry is restored when the inelastic
backscattering is switched off, and the turnstile character of the ring is removed when the
energy transfer to the ring is excluded for both transferred and backscattered electron waves.
We discuss the relation of the present results to the conductance systems based on the electron gas.
\end{abstract} \maketitle

\section{Introduction}
During two past decades a significant progress in control and manipulation of separate
electrons within the solid state devices has been made.
A single electron was trapped within a quantum dot \cite{tarucha} in a localized state.
Monitoring the flow of current by resolving the passage of separate electrons
has been achieved. \cite{onthefly} An ultrafast single-electron pumping in a system of quantum dots
connected in series was realized.\cite{pump}
Single-electron Aharonov-Bohm interference was demonstrated \cite{gustav} using a Coulomb-blockaded quantum dot as a valve
injecting separate carriers into the channel via cotunneling events.
Recently, single-electron transfer in a channel placed above the Fermi energy of the reservoirs
was reported \cite{ondemand} with the surface acoustic waves used to trap the moving carrier.

A single electron moving within
the channel  can be scattered inelastically and pass its energy to
the environment. On the other hand for the conventional experiments with the electron gas, inelastic scattering of the Fermi level electrons is forbidden
by the Pauli exclusion principle. The electron transport is strictly a Fermi level property in the linear
regime, where the current $I$ is necessarily an even function of the external magnetic field $B$, i.e. $I(B,V)=G(B)V$,
where $G$ is the linear conductance and $V$ the applied bias. The Landauer-B\"uttiker approach
derives the linear conductance $G(B)=\frac{e^2}{h}T(B)$ out of the electron transfer probability $T$,
and the latter is an even function of the magnetic field $T(B)=T(-B)$. The Onsager-Casimir \cite{OCo} symmetry $G(B)=G(-B)$  \cite{OC} does not hold for the non-linear transport,\cite{mb} where a finite energy window participates in the
current flow. Asymmetry of conductance by the non-linear currents carried by the { electron gas} was studied both experimentally\cite{glja,wei,let,zum,ch,bk,na} and theoretically\cite{mb,sz,bs,dsz,pb,dm,ag,sk,ar,li} in a number of papers.

Here we consider { a single} electron injected into a quantum wire and its probability
to pass through an interaction range of another electron confined in
a quantum ring placed in neighborhood, close enough to allow the capacitive coupling \cite{onthefly,gustav,ondemand} between the carriers.
We find that this probability is asymmetric in $B$.
We investigate the relation of the magnetic asymmetry
with the inelastic scattering effects.
We indicate that the magnetic symmetry of the electron transfer is restored when the inelastic
backscattering is excluded. The latter is achieved by inserting a narrow band-pass energy filter in form
of a double barrier structure into the channel
with the resonant energy fixed at the energy of the incoming electron. We show that the energy filter introduced
into the channel restores the magnetic symmetry of the transfer probability only for the electrons
traveling in one direction and not the other, hence the turnstile character of the system is observed
with or without the energy filter.

An appearance of the magnetic asymmetry of the single electron transfer probability was previously discussed
in an bent quantum wire \cite{kalina} or a cavity \cite{szafran} asymmetrically connected to terminals.
 Both papers \cite{kalina,szafran} used
a time dependent wave-packet approaches and indicated that the asymmetry of the transfer probability arises when the
channel electron interacts with the surrounding environment.
 The present study of the role
of the inelastic scattering requires a discussion of the incoming electron of a definite energy rather than the wave packet dynamics.
We develop such an approach
below and explain its relation to wave packet scattering. The results of this paper are based on a solution
of the two-electron Hamiltonian eigenequation with an exact account taken for the interaction and the electron-electron correlations.

This paper is organized as follows.
In the next section we first sketch the two-electron Hamiltonian used in this paper
in strictly one dimensional models of both the wire and the ring.
Next, we present a time-dependent approach to the scattering problem and then the time-independent treatment.
We demonstrate that the results of the latter can be understood as the limit
of monoenergetic wave packet scattering.
Section III contains the results and IV the discussion.  Summary and conclusions are given in section V.

  \begin{figure}[ht!]
     \centering

     \hbox{\rotatebox{0}{
                    \includegraphics[bb=0 0 230 360,  width=50mm] {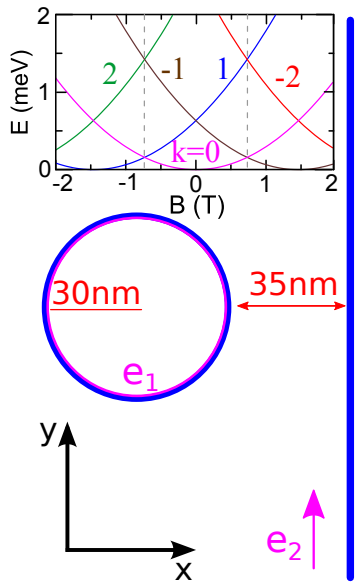}
                    }}
                     \caption{Schematics of the considered system: the electron $e_2$ travels along a straight channel and is scattered
                     on the potential of $e_1$ electron that is confined in a quantum ring of radius 30 nm placed at a distance of 35 nm from the channel. The top plot shows the energy spectrum of the electron in the ring.}
                     \label{schemat}
      \end{figure}

\section{Theory}
The system considered in this paper is schematically depicted in Fig. \ref{schemat}.
An electron is confined in a circular quantum ring of radius $R=30$ nm.
Initially, this  electron is in its ground-state, with a definite angular momentum
and circularly symmetric charge distribution. Another electron injected from outside goes along the straight channel, interacts with the ring-confined-electron and is partially backscattered. The total energy of the two-electron system is a conserved quantity.  The incoming electron is scattered inelastically when the ring absorbs a part of its energy.

The Hamiltonian of the electron in the circular ring with center
in point $(x_c,y_c,0)$ is given by $h_r=\frac{1}{2m^*}({\bf p}+e{\bf A})^2+V(r_c)$
with $r_c^2=(x-x_c)^2+(y-y_c)^2$. The magnetic field $(0,0,B)$ is oriented perpendicular to the plane of electron confinement. For the symmetric gauge
${\bf A}_s=\frac{B}{2}(-(y-yc),x-x_c,0)$ the Hamiltonian of the ring electron takes the form
$h_r=-\frac{\hbar ^2}{2m^*}\nabla^2+V(r_c)+\frac{e^2B^2}{8m^*}r_c^2+\frac{eB}{2m^*}l_c$,
where $l_c$ is the operator of the angular momentum $z$-component with respect to the ring center.
Operators $h_r$ and $l_c$ have common eigenstates $\phi^c_l=f_l(r_c)\exp(il\theta)$, with the angular momentum quantum number $l$. In the limit of a thin ring
the radial wave function $f_l$ tends to the ground-state of a particle confined in an infinite quantum well
and looses its dependence on $l$. The energy spectrum is then given by
$\varepsilon_l=E_r+\frac{\hbar^2}{2m^*R^2}(l+\frac{\Phi}{\Phi_0})^2$ (see the inset to Fig. 1), where
$\Phi_0=\frac{h}{e}$ is the flux quantum, $\Phi=B\pi R^2$ and $E_r$ is the ground-state energy of the radial confinement.
The latter is independent of $l$ and as such is irrelevant for the scattering process. We skip $E_r$ in the following formulae.

For the scattering problem it is most convenient to use another gauge ${\bf A}=B(0,x,0)$, since
then the diamagnetic term produced by the kinetic energy operator ($\frac{e^2B^2}{8m^*}x^2$) vanishes
at the axis of the channel $x=0$. In the following we assume that the channel is so thin that the  electron in its
motion along the channel
is in its lowest state of lateral quantization.
For the strictly 1D channel with $x=0$ axis the kinetic momentum  $\pi_y={\bf p}_y+eBx$
is independent of $B$, and thus  the wave vector $q$ of the motion along the lead corresponds to the same energy
and probability current flux for any $B$.

In order to replace ${\bf A}_s$ by ${\bf A}$ the gauge transformation ${\bf A}={\bf A}_s+\nabla \chi(x,y)$ is performed with $\chi=\frac{B}{2} (xy+x_cy-y_cx)$. Upon the transformation the ring wave functions change to \begin{equation} \phi_l=\phi^c_l(r_c,\theta) \exp(-\frac{ie}{\hbar}\chi(x,y)), \label{gt}\end{equation}
where the phase factor introduced by $\chi$ is independent of $l$.
Although with {\bf A} the angular momentum with respect to the ring center does not commute with the Hamiltonian, $l$
still remains a good quantum number for description of the ring eingestates.

With the assumptions explained above the two-electron Hamiltonian used in this work reads
\begin{equation}
H=h_c({\bf r}_1)+h_r({\bf r}_2)+W(|{\bf r}_1-{\bf r}_2|), \label{ww}
\end{equation}
where $h_c=-\frac{\hbar^2}{2m^*}\frac{\partial^2}{\partial y^2}$ is the channel electron Hamiltonian and $W$ is the interaction
potential. The latter is taken in the screened Coulomb form
\begin{equation}
W(r)=\frac{e^2}{4\pi\epsilon \epsilon_0 r}\exp(-r/\lambda),
\end{equation}
with dielectric constant  $\epsilon=12.9$ and the screening length $\lambda=500$ nm

\subsection{Time-dependent scattering picture}
The general form of the two-electron wave function can without a loss of generality be developed in the basis of product
of single-particle eigenstates with definite angular momentum for the ring and the wave vector within the channel $q$
\begin{eqnarray}
\Psi({\bf r}_1,{\bf r}_2,t)&=&\sum_{ql} c_{ql} (t) \Phi_q({\bf r}_1)\phi_l({\bf r}_2) \label{pw}\\ &=& \sum_l \psi_l({\bf r}_1,t)\phi_l({\bf r}_2),\label{dw}
\end{eqnarray}
where the partial wave packets are defined as
\begin{equation} \psi_l({\bf r}_1,t)\equiv\sum_q c_{ql}(t)\Phi_q({\bf r}_1).\label{aas} \end{equation}
The electrons occupying separate regions in space (the wire and the ring) are essentially distinguishable. Anti-symmetrization
of Eq. (\ref{aas}) does not affect any of the results presented below due to the complete separability of the electron wave functions.\cite{dudziak}
For that reason we skipped the anti-symmetrization in the following.

One puts the wave function (\ref{dw}) into the Schr\"odinger equation $i\hbar \frac{\partial \Psi}{\partial t}=H\Psi$
and projects the result on the ring eigenstates, which leads to a set of equations for the partial wave packets
\begin{equation}
i\hbar \frac{\partial \psi_k(y_1,t)}{\partial t}=\sum_l \left([\varepsilon_l+h_c] \delta(k,l)+W_{kl}(y_1)\right)\psi_l(y_1,t),
\end{equation}
where $W_{kl}({\bf r}_1)=\langle \phi_k ({\bf r}_2)|W (|{\bf r}_1-{\bf r}_2|)| \phi_l({\bf r}_2)\rangle$.
Note, that the phase factor due to the gauge transformation ($\ref{gt}$) is canceled in the evaluation of the interaction matrix $W_{kl}$.

In the time dependent calculation we take for the initial condition a Gaussian wave packet $\Psi_l(y,t)=\sqrt{\frac{\Delta k}{2\pi^{1/4}}}\exp(-\frac{\Delta k^2}{4}(y-y_0)^2+iqy)$,
where $l$ corresponds to the ground-state angular quantum number, the average momentum $q>0$ and $y_0$ is far below the ring. For $k\neq l$, in the initial condition $\Psi_k=0$ is applied.
Calculations are performed with a finite difference scheme for the channel of length 16 $\mu m$ with $\Delta y=2$ nm.
The results converge when $|l|\leq 3$ ring eigenstates are included into the basis.

\subsection{Stationary description of the scattering}
The time-independent approach described in this section is suitable for treating the scattering for the incident electron of a definite energy.
The stationary approach is also more computationally effective and does not require
very large computational box since transparent boundary conditions can readily be applied.  For $\Delta k=0$ the incoming electron has a definite momentum $\hbar q$,
and a definite energy $E_i=\frac{\hbar^2 q^2}{2m^*}$, hence the total energy $E_{tot}$ of the system is also a well-definite quantity $E_{tot}=E_i+\varepsilon_l$, where $\varepsilon_l$ is the ring ground-state energy.
Therefore, the two-electron wave function for the scattering satisfies the time-independent Schr\"odinger equation
\begin{equation}
H \Psi({\bf r}_1,{\bf r}_2)=E_{tot}\Psi({\bf r}_1,{\bf r}_2). \label{ti}
\end{equation}
We use the form of the function
\begin{equation}
\Psi({\bf r}_1,{\bf r}_2)=\sum_l \psi_l({\bf r}_1)\phi_l({\bf r}_2),\label{tidw}
\end{equation}
which is a time-independent counterpart of Eq. (\ref{dw}).
Insertion of Eq. (\ref{tidw}) into Eq. (\ref{ti}) followed by projection on a ring eigenstate gives
a system of eigenequations for $\psi_l$,
\begin{equation}
\sum_l \left([\varepsilon_l +h_c]\delta(k,l)+W_{kl}(y)\right)\psi_l(y)=E_{tot} \psi_k(y). \label{eqs}
\end{equation}

The electron in the ring is initially in its ground-state with angular momentum $l$ -- as in the
time independent picture.
Therefore, the partial wave $\psi_l$ at the input side is a superposition
of the incoming and backscattered waves $a\exp(i q_l y)+b\exp(-iq_ly)$. Since $\Psi$ is defined
up to a normalization constant, at the bottom of the computational box ($3\mu m$ long)
we simply set $\psi_l(0,y=0)=a+b=1$ as the boundary condition. After the solution
of Eqs. (\ref{eqs}) the values of the incoming $a$ and the backscattered $b$ amplitudes
are extracted from the form  $\psi_l$ along the lead.

The partial waves for $k\neq l$ appear only due to the interaction of the incoming electron with the ring,
and they all correspond to the electron flow from the ring to the ends of the channels.
Thus,  far away above [below] the ring the partial wave functions corresponding to $k$-th angular momentum quantum number correspond
to transferred [backscattered] electron and have the form of $c_k \exp(i q_k y)$
[$d_k \exp(-i q_k y)$], with $q_k=\sqrt{\frac{2m^*}{\hbar^2}\left(E_{tot}-\varepsilon_k\right)}$.
For $E_{tot}>\varepsilon$ the wave vector $q_k$ is real and the boundary condition $\psi_k(y+\Delta y)=\psi_k(y) \exp(i q_k \Delta y)$
[$\psi_k(y+\Delta y)=\psi_k(y) \exp(-i q_k \Delta y)$] is applied at the top [bottom] end of the computational channel.
For $E_{tot}<\varepsilon$ the wave vector $q_k$ is imaginary and the wave function vanishes
exponentially along the lead. The partial waves with imaginary $q_k$ are counterparts of the evanescent modes \cite{EM} for scattering
in two-dimensional channels. For imaginary wave vectors we put zero for $\psi_k$ at the ends of the computational box.
Upon solution of Eq. ($\ref{eqs}$), the amplitudes $a,b,c_l,d_l$ are calculated. The total transfer probability
is given by $T=\sum_k T_k$ with $T_k=\frac{|c_k|^2}{|a|^2} \frac{q_k}{q_l}$, similarly
the backscattering probability is $R=\sum_k R_k$  with $R_k=\frac{|d_k|^2}{|a|^2} \frac{q_k}{q_l}$.

  \begin{figure}[ht!]
     \centering
    \begin{tabular}{l}

           \hbox{\rotatebox{0}{
                    \includegraphics[bb=100 250 600 800,  width=50mm] {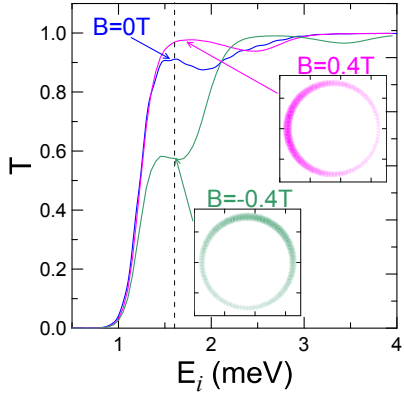}
          }}

 	\end{tabular}

    \caption{Transfer probability of electron transport through channel in function of the incoming electron energy for
    three values of the magnetic field. The inset shows the charge density as obtained by the stationary transport description
    for three values of the magnetic field.}
    \label{model}
    \end{figure}

  \begin{figure}[ht!]
     \centering

    \hbox{\rotatebox{0}{
                    \includegraphics[bb=100 250 400 800,  width=30mm] {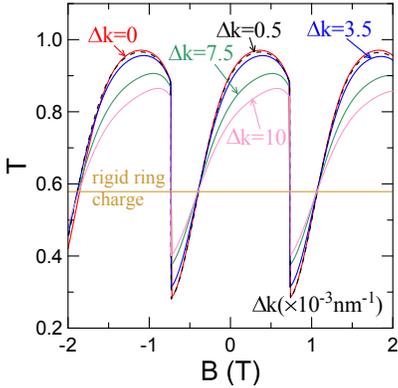}
                    }}

    \caption{The green, blue and black lines show the packet transfer probability through the system of Fig. 1
    as calculated by the wave packet simulation for a number of wave vector dispersions for $\Delta k\leq 10^{-3}$/nm as functions
    of the external magnetic field.  The red line shows the result of the time-independent scattering problem ($\Delta k=0$).
    The horizontal line shows the transfer probability for fixed charge of the ring.
     }
    \label{model}
    \end{figure}

 \begin{figure}[ht!]
     \centering
    \begin{tabular}{l}

(a)    \hbox{\rotatebox{0}{
                    \includegraphics[bb=100 250 600 780,  width=50mm] {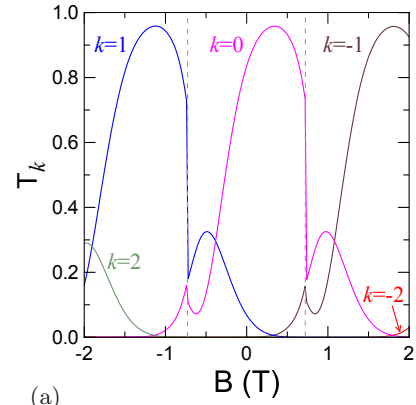} }}
                    \\ (b)
    \hbox{\rotatebox{0}{
                    \includegraphics[bb=100 250 600 780,  width=50mm] {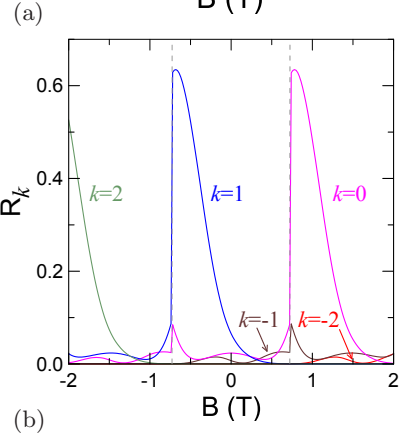} }}

 	\end{tabular}

    \caption{Transfer (a) and backscattering (b) probability associated with angular momentum $k$ of the ring
    in the final scattering process (see text).}
    \label{cnt}
    \end{figure}

  \begin{figure}[ht!]
     \centering

     \begin{tabular}{l}
     \hbox{\rotatebox{0}{
                    \includegraphics[bb=200 250 350 600,  width=15mm] {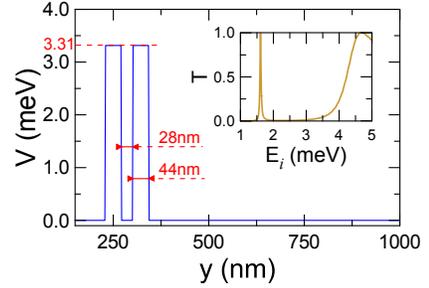}
                    }}
     \end{tabular}
      \caption{  Double barrier structure used a the energy filter. The inset shows transmission probability through the barrier in function energy
      with a peak at $1.6$ meV -- the incident electron energy. The ring center is set at $y_c=1500$ nm. }
      \end{figure}

  \begin{figure}[ht!]
     \centering

     \begin{tabular}{l}
     \hbox{\rotatebox{0}{
                    \includegraphics[bb=200 250 350 700,  width=20mm] {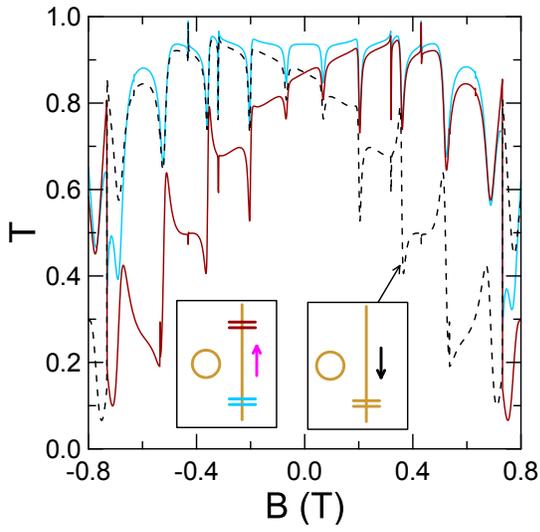}
                    }}
     \end{tabular}
      \caption{Electron transfer probability for the DBS placed below (red curve)
      or above (blue curve) the ring for the electron incident from the lower end of the wire.
      The dashed curve shows the transfer probability for the electron going down with the
      DBS placed below the ring.
        }
      \end{figure}

  \begin{figure}[ht!]
     \centering

     \begin{tabular}{l}
     \hbox{\rotatebox{0}{
                    \includegraphics[bb=200 250 350 600,  width=20mm] {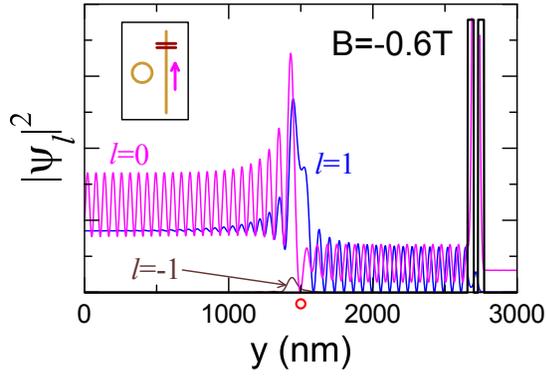}
                    }}
     \end{tabular}
      \caption{Density of partial waves for the DBS placed above the ring. Location of the ring ($y=1500$ nm) is marked
      by a circle on the horizontal axis.}\label{xox}
      \end{figure}

  \begin{figure}[ht!]
     \centering

     \begin{tabular}{l}
      (a)
     \hbox{\rotatebox{0}{
                    \includegraphics[bb=100 250 600 800,  width=60mm] {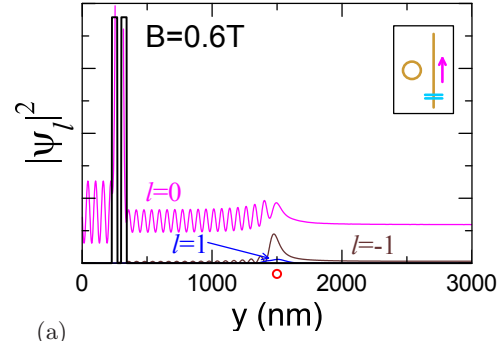}
                    }} \\ (b)
      \hbox{\rotatebox{0}{
                    \includegraphics[bb=100 250 600 800,  width=60mm] {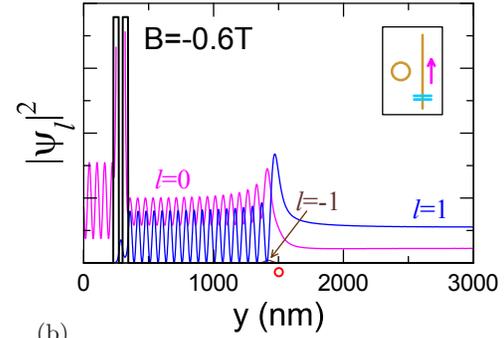}
                    }}
     \end{tabular}
      \caption{ Same as Fig. \ref{xox}, only for the DBS placed below the ring. }
      \end{figure}

  \begin{figure}[ht!]
     \centering

     \begin{tabular}{l}
     \hbox{\rotatebox{0}{
                    \includegraphics[bb=200 250 350 700,  width=15mm] {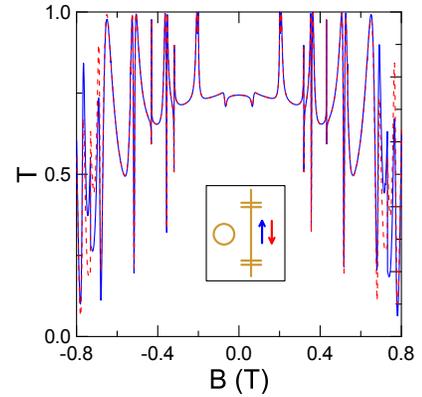}
                    }}
     \end{tabular}
      \caption{Electron transfer probability for the DBS placed both below
      and above the ring for the electron going up (blue curve) or down (red dashed curve) the wire.}
      \end{figure}

  \begin{figure}[ht!]
     \centering

     \begin{tabular}{l}
     \hbox{\rotatebox{0}{
                    \includegraphics[bb=200 250 350 700,  width=17mm] {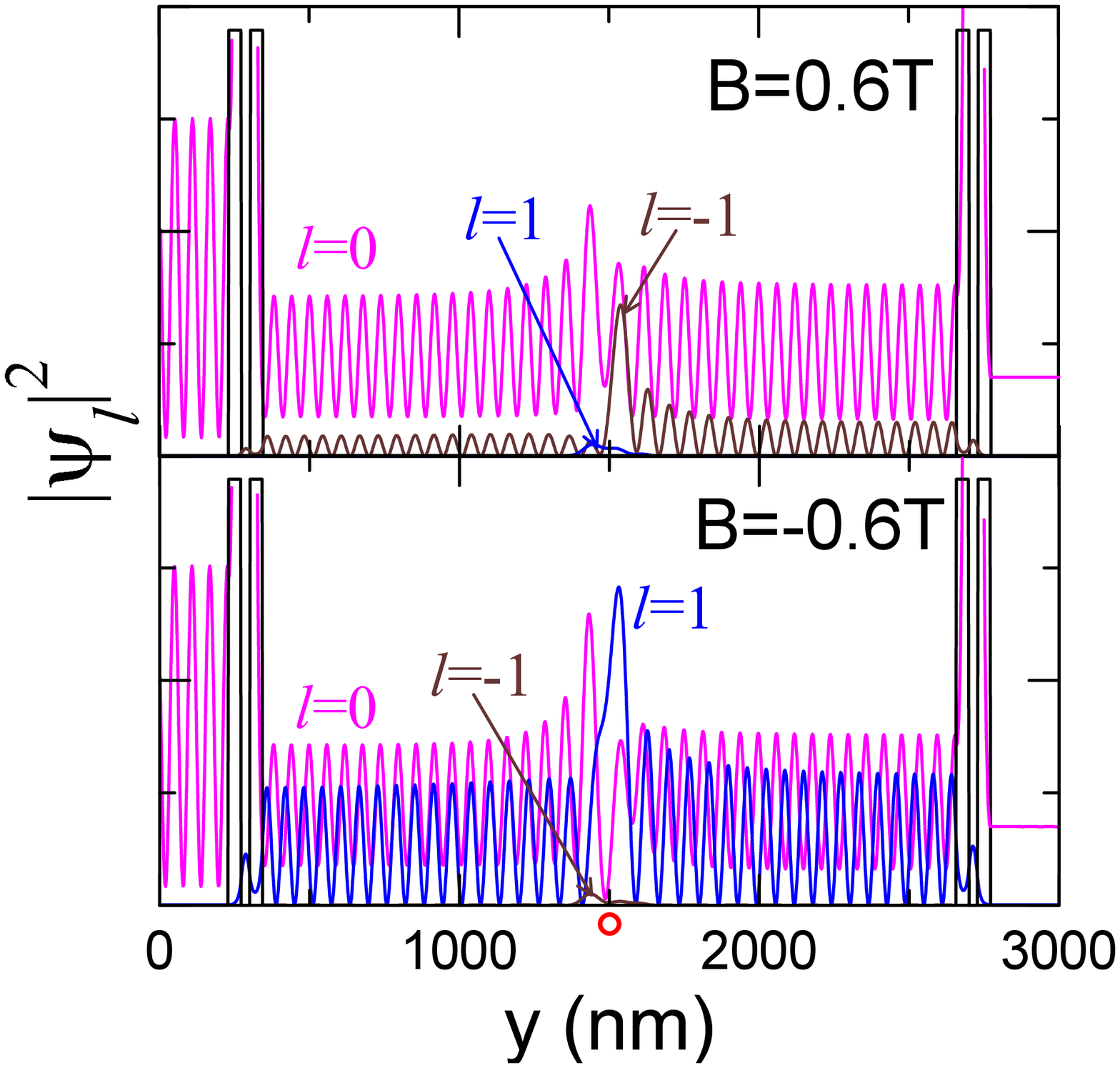}
                    }}
     \end{tabular}
      \caption{ Same as Fig. \ref{xox} only for two DBS: one below and second above the ring. }
      \end{figure}

\section{Results}

In Fig. 2 we plotted the electron transfer probability obtained by the time-independent method in function of the incident electron energy,
for three values of the magnetic field. For $E_i<1$ meV the transfer probability vanishes
and for $E_i>3$ meV the value of $T$ becomes close to 1 independent of $B$. Around $E_i=1.6$ meV a distinct asymmetry of $T$ as a function of $B$ is found.
The insets displays the charge density within the ring calculated as $\rho({\bf r}_2)=\int d{\bf r}_1 |\Psi({\bf r}_1,{\bf r}_2)|^2$.
For $B=0.4$ T the density is shifted off the channel (at right to the ring), and consistently $T$ is larger. 

The results of the time-dependent simulation for the packet {\it average} energy of $\langle E_i\rangle=\frac{\hbar ^2 q^2}{2m^*}=1.6$ meV are plotted in Fig. 3 in function of $B$ for a number of initial dispersions of the wave vector $\Delta k$.
The horizontal line shows the result obtained for a rigid charge of the ring  which is independent of $B$.
All the $B$ dependence of the transfer probabilities given in Fig. 3 is due to the properties of the ring as an inelastic scatterer
which change with the magnetic field. The discontinuities present in the transfer probabilities at $B=\pm B_0=\pm 0.73$ T result from ground-state angular
momentum transitions within the ring [see the top inset to Fig. 1]. With $\Delta k$ decreasing to 0 the results converge to the
 result of the stationary description of the scattering for the
incoming electron energy of {\it definite} energy $E_i=\frac{\hbar ^2 q^2}{2m^*}=1.6$ meV which are plotted with
the red line in Fig. 2.
The rest of the results presented in this work was obtained with the stationary description of the transport.

The electron transfer probability as depicted in Fig. 3 is a distinctly asymmetric  function of $B$. The asymmetry along with the character of the discontinuities
at the ring ground-state transformations can be understood as due to the relation of the backscattering to the angular momentum absorption by the ring.  The incoming (backscattered) electron has a positive (negative) angular momentum with
respect to the center of the ring. When the ring electron compensates for the loss of angular momentum the backscattering is more probable.
Let us concentrate on the magnetic field interval $[-B_0,B_0 ]$ in which the ring ground state corresponds to $l=0$.
The absorption of the angular momentum by the ring is associated with transition from $l=0$ to $l=1$ energy level.
This is less energetically expensive when $B$ becomes negative due to decreasing energy spacing
between the ground state energy and the $l=1$ energy level (see the inset to Fig. 1).
Consistently, the contribution of $l=1$ energy level to the total backscattering probability grows
 as $B$ decreases below 0 -- see Fig. 4(b). Fig. 4(a) shows that for $B$ just above the ring-state transition $l=1$ ring state dominates also in the transfer probability.

Below the ground-state  angular momentum transition which occurs at $B=-0.73$ T
the ring ground state $l$ is 1 and the absorption of angular momentum by the ring requires an appearance
of $l=2$ wave function to the final scattering process. This becomes energetically expensive below $B<-B_0$,
hence the jump of $T$ that is observed in Fig. 3 at the ring ground-state transformation. As $B$ is decreased further $T$ drops and $l=2$ starts to dominate in the backscattering probability [see Fig. 4(b)].

Our results for the single-electron scattering indicate that the energy absorption is associated both with the electron transfer [Fig. 4(a)] and backscattering [Fig. 4(b)], which is accompanied by magnetic symmetry violation for the electron transfer probability.
We found that one can eliminate selectively the effects of inelastic scattering in the transferred or backscattering waves by a proper tailoring of the potential profile along the channel.
For that purpose we used a double barrier structure (DBS) with center placed on the channel far (1200 nm) below the ring.
Figure 5 shows the applied potential profile and the inset to the figure the electron transfer probability
through the DBS. We can see the resonant peak at the electron energy of 1.6 meV.
 The resonant energy was set equal to the energy of the incoming electron, so that the DBS acts like an energy filter -- it is opaque for the electron that lost a part of its energy, i.e. to the partial waves with $k\neq l$.

In Fig. 6 we plotted with the red line the transfer probability for the DBS energy filter placed above the ring.
Fig. 7 shows the plot of partial waves along the channel. Above the DBS
one finds only the partial wave associated with $l=0$, i.e. with the ground-state of the ring.
The electron can transfer across the structure only provided that the it preserves its initial energy.
Therefore, no excitation of the ring electron is possible when the channel electron transfers across the structure.
In Fig. 7 we can see that far below the ring we have an interference of  $l=0$ incoming and backscattered waves.
No interference is observed in the partial wave with $l=1$ near $x=0$ ($|\psi_1|$ is constant), since
there is no incoming wave with $l=1$. Nevertheless an oscillation of $l=1$ wave is observed between the
DBS  and the ring. The potential of the ring and the DBS form
a wide quantum well in which the  partial waves [for instance $l=1$ in Fig. 7] oscillate back and forth. The presence of the wide well
is also responsible for the resonances appearing at the $T(B)$ dependence in Fig. 6. $T(B)$ for the DBS placed above the ring remains an asymmetric function of $B$.

The transfer probability $T$ becomes an  even function of $B$ (blue curve in Fig. 6) when the DBS energy filter
is placed below the ring, which removes inelastically scattered partial waves of the total backscattered wave function.
The partial wave function plots given in Fig. 8(a) and Fig. 8(b) show that below the DBS only the partial wave with $l=0$ is found,
but above the structure we see an appearance of the partial waves for $l\neq 0$.

For $B>0$ just below $B_0$ we found that $T(B)$ is nearly the same for the double barrier structure
placed both below and above the ring [see the blue and red curves which nearly coincide in Fig. 5 just below $B_0$].
Note, that for the DBS below the ring at $B=0.6$ T we find that the contribution of $l\neq 0$ in the transferred wave function
is negligible [Fig. 8(b)]. The absorption of the  angular momentum by the ring is weak for $B\rightarrow B_0$ due to the large
energy cost of this ring excitation [see the discussion of Fig. 2], hence the similar results found for both locations
of DBS.

In Fig. 6 with the dashed curve we plotted the electron transfer probability for the DBS below the ring
and the electron incident from the upper end of the wire. In this case the electron is first scattered by
the ring and then by the DBS. We can see that for a single DBS present within the wire the transfer probability from one
end of the wire to the other is different than in the opposite direction (the dashed curve in Fig. 6 can be obtained from
the red one by inversion $B\rightarrow -B$), i.e. the system acts like a turnstile.

Figure 9 gives the electron transfer probability for two DBS placed both below and above the ring.
The inelastic scattering is switched off for both the transferred and backscattered trajectories.
The partial waves given in Fig. 10 show that the ring does get excited but only for the channel
electron staying between the two DBS.
We find that the transfer probability is symmetric with respect to both the magnetic field and the direction
from which the electron comes to the ring. The small deviations off the symmetries visible at a closer inspection of Fig. 9
 are due to small but finite width of the resonance peak (see the inset to Fig. 5).
The inelastic scattering is allowed with the energy losses smaller than the width of the peak.

\section{discussion}

Results of Figs. 3 and 4 indicate that the asymmetry of the transfer probability as a function of the magnetic field
is a result of 1) geometrical asymmetry of the system 2) inelastic electron scattering
 -- the absorption of the angular momentum by the ring
which is necessarily accompanied by the energy absorption 3) the energy transfer occurs through the electron-electron interaction.

For systems with the two-dimensional electron gas
it was pointed out \cite{mb,bs} that  the magnetic asymmetry of conductance may result from the potential
landscape within the device being not an even function of $B$ -- the potential produced by charges
at the edges of the channel in the Hall effect \cite{mb} as the most basic example. In this case
the asymmetry of the charge distribution is translated to the asymmetry of the transport by the electron-electron
interaction.
The role of the electron-electron interaction for the magnetic asymmetry of the transport in the electron gas was also indicated in Refs. \cite{bs,dsz,sk}.
In the present study of the single-electron transport the asymmetry is due to the properties of the ring -- the enhancement
of the backscattering accompanied by absorption of the angular momentum of the channel electron -- which are not
an even function of $B$ due to the form of the ring energy spectrum. Here, the backscattering is only due to the electron-electron interaction.
Although in the linear transport regime the inelastic scattering of the electrons at the Fermi level is blocked by
the fact that the states of lower energies are occupied, in the non-linear transport the inelastic scattering
is not only allowed but necessary for thermalization of the carriers passing between electron reservoirs of
unequal electrochemical potentials. The asymmetry that we find in this work results from the energy transferred
by the channel electron to the ring, i.e. it occurs due to the inelastic scattering.
The magnetic symmetry is restored when the inelastic backscattering is excluded. The invariance of the backscattering
is invoked in explaining $T(B)=T(-B)$ symmetry when
the transfer kinetics is very different for both magnetic field orientations -- see
the deflection of the electron trajectories by the Lorentz force in Ref. \onlinecite{kalina}.
In the present work the Lorentz force is excluded by the strict 1D approximation for the channels width.
Nevertheless, the different kinetics resulting in the same transfer probability was also found
in Figs. 8(a) and 8(b).

A single DBS placed below the ring restores the magnetic symmetry of the transfer, still only
for the electron injected from one side of the channel and not the other  (the microreversibility is not restored -- see Fig. 6).
Thus, for a single DBS present within the wire the electron transfer probability from one end of the terminal to the other
are unequal. The turnstile character of the system is also a result of the inelastic scattering.
The conditions present in the linear transport regime -- with the inelastic scattering
excluded at both the transfer and the backscattering -- were simulated with two DBS placed
at both sides of the ring. This configuration of energy filters
restores the microreversibility of the system.
The transfer probability becomes
an even function of $B$,
although the  kinetics of the electron transfer is not identical for $\pm B$ [Fig. 10].
Moreover, the microreversibility is also restored [Fig. 9], although
the system  with two DBS is still not spatially symmetric under a point inversion.

\section{Summary and Conclusions}

We have studied single-electron scattering process on an electron localized in a quantum ring off the electron transport channel.
We developed for that purpose a time-independent approach based on an expansion of the two-electron function in a basis
of ring eigenstates and explained its relation to the numerically exact time-dependent scattering picture.
We have found that the electron transfer probability is an asymmetric function of $B$ and that the asymmetry
results from the energy cost of the angular momentum absorption by the ring which is not an even function of $B$.
We have demonstrated that the symmetry is restored when the electron backscattering with the energy loss is excluded.
The exclusion was performed by a double barrier structure with the resonant state set at the energy of the incoming electron.
In order to remove the turnstile character of the ring as a scatterer one needs to employ a pair of double barrier
structures at both the entrance and the exit to the ring interaction range.

    {\bf Acknowledgements}
This work was performed
supported by the Polish Ministry of Science an Higher Education (MNiSW) within a research project N N202 103938 for 2010-2013.
 Calculations were  performed in    ACK\---CY\-F\-RO\-NET\---AGH on the RackServer Zeus.

\end{document}